\documentclass{qjrms2}

\usepackage[mathscr]{eucal}	   

\Times 
\begin{document}

\QJRMS{1}{999}{128}{2002}{yy.n}
\runningheads{C. D. WESTBROOK \textit{et al.}}{RADAR SCATTERING BY AGGREGATE SNOWFLAKES}
\title{Radar scattering by aggregate snowflakes}
\author{C. D. WESTBROOK$^1$, R. C. BALL$^1$ and P. R. FIELD$^2$\footnote{Corresponding author: National Centre for Atmospheric Research, Boulder, Colorado, USA; e-mail prfield@ucar.edu}}
\affiliation{$^1$University of Warwick, UK\\
$^2$National Centre for Atmospheric Research, USA.} 
\date{(Received 1 January 2000; revised 31 January 2001)}
\rmscopyright 
\begin{abstract}

The radar scattering properties of realistic aggregate snowflakes have been calculated using the Rayleigh-Gans theory. We find that the effect of the snowflake geometry on the scattering may be described in terms of a single universal function, which depends only on the overall shape of the aggregate and not the geometry or size of the pristine ice crystals which compose the flake. This function is well approximated by a simple analytic expression at small sizes; for larger snowflakes we fit a curve to our numerical data.
We then demonstrate how this allows a characteristic snowflake radius to be derived from dual-wavelength radar measurements without knowledge of the pristine crystal size or habit, while at the same time showing that this detail is crucial to using such data to estimate ice water content. We also show that the `effective radius', characterising the ratio of particle volume to projected area, cannot be inferred from dual-wavelength radar data for aggregates. Finally, we consider the errors involved in approximating snowflakes by `air-ice spheres', and show that for small enough aggregates the predicted dual wavelength ratio typically agrees to within a few percent, provided some care is taken in choosing the radius of the sphere and the dielectric constant of the air-ice mixture; at larger sizes the radar becomes more sensitive to particle shape, and the errors associated with the sphere model are found to increase accordingly.

\end{abstract}
\ahead{Introduction} 
            %
	    
	    Aggregation plays an important role in the evolution of ice clouds and the development of precipitation. Understanding the geometry and size distribution of ice aggregates is crucial to analysing radar returns and interpreting them in terms of particle size and ice water content. 
	    A recent study (Westbrook \textit{et al} 2004a,b) presented a new theoretical model of ice crystal aggregation which produced `synthetic' snowflakes with realistic geometry and distribution by size; sample aggregates from those simulations are shown in figure 1, alongside images of real aggregates from a cirrus cloud. This model includes no \textit{a priori} assumptions about the relationship between snowflake mass and dimension, or about the distribution by size. The results from this study show that many of the features of the resulting aggregate snowflakes are \textit{universal} -- that is to say that they do not dependend on the shape or initial size distribution of the pristine ice crystals which make up the flakes.

	    The purpose of this paper is to calculate the radar scattering properties of these synthetic aggregates and to consider how the results may be used to interpret dual wavelength radar data in terms of average particle size and ice water content. We identify a number of features in the scattering behaviour of the snowflakes which are universal and, equally importantly, a number of features which are not. These results have significant implications for the estimation of the microphysical properties of ice clouds from radar data.
	    We also consider the errors involved in modelling the aggregates as spherical ice/air mixtures, and in particular what size sphere should be used in place of the aggregate it is intended to represent.  
	    \begin{figure}                                          
            \centering\includegraphics{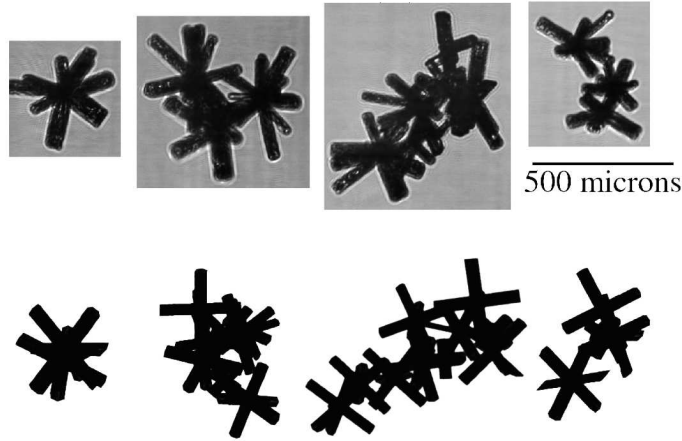}             
            \caption{\label{simandcpi} Aggregates of bullet-rosette crystal types: top half shows images obtained from an aircraft flight through a cirrus cloud at temperatures between $-44^{\circ}$C and $-47^{\circ}$C ($\sim 9$km altitude), using a cloud particle imager. Lower half shows projected images of some of the `synthetic' aggregates produced by our computer simulations (see text).}                                
            \end{figure}

\ahead{Aggregation model}
Here we present a brief overview of the Westbrook \textit{et al} (2004a,b) model, and the key results from it. The model is based on the rate of hydrodynamic capture between two spheres, well known in the raindrop coalescence literature:
\begin{equation}
\Gamma_{ij}=\frac{\pi}{4}(D_i+D_j)^2\left|v_i-v_j\right|,
\end{equation}
where $D$ is the maximum particle dimension and $v$ is the fall speed. Unlike raindrops, snowflakes are clearly not spherical: in order to accurately sample collisions between the complicated aggregate shapes we use $\Gamma_{ij}$ as a rate of `close approach' between the two particles $i$ and $j$. We pick pairs of particles at random with a probability proportional to $\Gamma_{ij}$, so as to choose two that are likely to undergo a collision. We then track the pair along one of the possible trajectories that the collision area $\frac{\pi}{4}(D_i+D_j)^2$ encompasses. If a collision occurs, the particles are stuck together rigidly at the point of initial contact; if the pair miss one another, they are returned to the system and a new pair is picked. In this way we correctly sample the collision rates between the complex aggregates.
 
The model is completed by an explicit form for the particle fall speeds, and this is provided by Mitchell (1996):
\begin{equation}
v\propto\frac{\nu_k}{r}\left(\frac{mg}{\rho\nu_k^2}\right)^{\alpha}
\label{fallspeeds}
\end{equation}
where $mg$ is the particle weight and $r$ is a characteristic radius (see below). The parameter $\alpha$ governs the hydrodynamic regime: here we assume an approximately inertial flow and set $\alpha=\frac{1}{2}$. The air enters the expression through its density $\rho$ and its kinematic viscosity $\nu_k$. The characteristic radius in equation \ref{fallspeeds} is the radius of gyration (see appendix A); however systematically replacing $r$ with the maximum dimension $D$ (as per Mitchell) does not affect the results described here. Our computer simulations show that $r\propto D$, and since we are only interested in how likely one collision is relative to another (in order to pick pairs of particles), it is unimportant which characteristic length scale we use to calculate $v$.
From the point of view of the radar scattering results the radius of gyration is a particularly natural length scale, and in what follows we will in general use $r$ rather than $D$. For reference, our simulation results show that $r\simeq0.3D$.

\ahead{Snowflake geometry and size distribution}

The aggregate snowflakes produced by the model have a fractal geometry (in a statistical sense). Consequently the flakes have a rather open structure, with a power law relationship between mass and radius:
\begin{equation}
m=ar^{d_f}.
\label{fractal}
\end{equation}
where the exponent $d_f$ is the fractal dimension and has value of less than three. This kind of power law scaling has been widely reported in the literature, usually with an exponent of around two (eg. Heymsfield \textit{et al} 2002: $d_f=2.04$ for aggregates of bullet-rosettes, $d_f=2.08$ for aggregates of side-planes; Locatelli and Hobbs 1974: $d_f=1.9$ for aggregates of plates, side-planes, bullets and columns; Mitchell 1996: $d_f=2.1$ for aggregates of side-planes, columns and bullets).
This tallies well with our simulation results where we measure $d_f=2.05\pm0.1$, and our theoretical arguments (see Westbrook \textit{et al} 2004b) which lead to the prediction $d_f=1/(1-\alpha)=2$. 

\begin{figure}                                          
            \centering\includegraphics{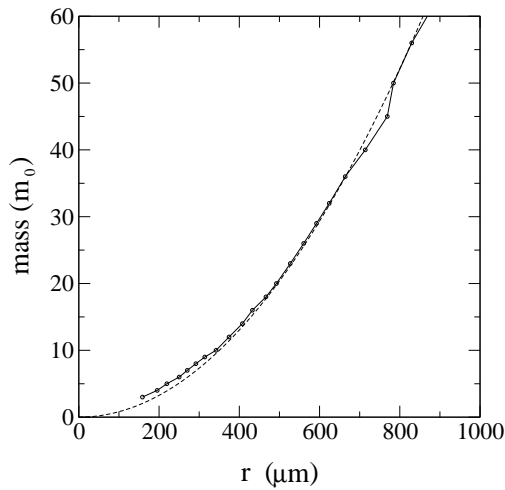}             
            \caption{\label{mr} Snowflake mass as a function of radius of gyration for synthetic aggregates of 50 and 100$\mu$m columns. Mass is in units of $m_0$ (the average mass of the pristine columns). The simulation data (circles) quickly approaches the asymptotic $m\propto r^2$ scaling (dashed line).}                           
            \end{figure}

Crucially this asymptotic scaling is determined purely by the hydrodynamic regime and not the initial conditions (pristine crystal type or size). These details do have an affect on how quickly the asymptotic regime is approached, but as figure \ref{mr} illustrates, in general only a few collisions are needed before equation \ref{fractal} applies.  
	    
	    Unlike $d_f$, the prefactor $a$ is not universal, and contains all the information relating to the (average) mass $m_0$ and radius $r_0$ of the pristine particles: $a=\gamma m_0/r_0^{d_f}$. The dimensionless geometrical factor $\gamma$ is determined by the crystal shape.
	    
	    The size distribution for our synthetic aggregates was found to approach a universal underlying shape, which spreads out as the aggregation continues:
\begin{equation}
n(m,t)=m_{av}^{-\xi}\phi(m/m_{av})
\label{ds}
\end{equation} 
where we define the weight average mass $m_{av}(t)=\sum m^2/\sum m$ and $\xi$ is a positive constant. The function $\phi$ describes the underlying distribution shape and its form is in principle sensitive only to the value of $\alpha$ in the velocity law (\ref{fallspeeds}).
As the aggregation progresses, the average mass $m_{av}$ increases, stretching the distribution whilst reducing the overall concentration through the factor $m_{av}^{-\xi}$. Equation \ref{ds} reflects the idea that there is a single underlying distribution which is rescaled depending on how far the aggregation has evolved (as characterised by the average aggregate mass $m_{av}$).

\begin{figure}                                          
            \centering\includegraphics{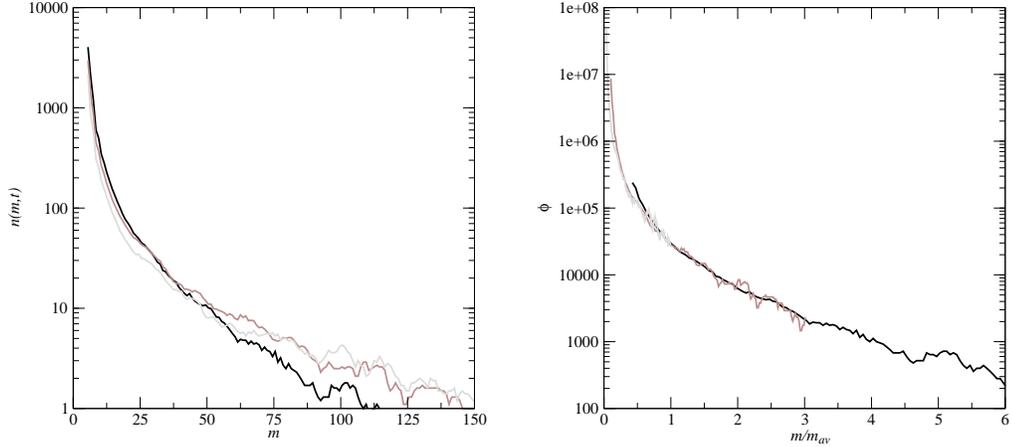}             
            \caption{\label{usands} Size distributions by mass for our synthetic snowflakes. Left panel shows snapshot distributions at three different times in the evolution of the size distribution: $m_{av}(t)=20$ (black line), $55$ (dark grey line) and $150$ (light grey line). Mass is in units of $m_0$ as per figure \ref{mr}; the simulation started with 5000 columns of unit mass, and 5000 with twice that weight. In the right hand panel we plot $m_{av}^{\xi}n(m,t)$ against $m/m_{av}$ to show that all three are in fact the same underlying distribution $\phi$ rescaled as a function of $m_{av}$. For our simulations total mass is conserved and as a result we take $\xi=2$.}                         
\end{figure} 

To test this `dynamical scaling' we measured $n(m,t)$ at three different points in the evolution of our simulations, and plot $m_{av}^{\xi}n(m,t)$ as a function of $m/m_{av}$ in each case. The results are shown in figure \ref{usands} and the data points from each of the three distributions collapse onto a single curve, confirming equation \ref{ds}. This result has also been confirmed using experimental size spectra from a cirrus cloud --- see Westbrook \textit{et al} (2004a,b) for details.  

	    In summary, the crucial results (backed up by experimental data) are that the relationship between mass and linear dimension takes a power law scaling, with a universal exponent determined by the physics of the aggregation process, and a non-universal prefactor which is sensitive to both the pristine crystal type and size. In addition, the aggregate size distribution is described by a universal underlying distribution function, which is simply rescaled as a function of the average particle mass as the aggregation proceeds.

\ahead{Radar cross sections of individual aggregates}		

\bhead{The Rayleigh-Gans theory} 
In this section the back scattered intensity from a single snowflake is calculated using the Rayleigh-Gans theory. 
The essence of this approximation is that the particle is split up into a number of small volume elements $\mathrm{d}v$ at position $\mathbf{r}$ relative to the overall centre of mass. Each element is treated as a Rayleigh scatterer, and we ignore any interactions between the elements (each one sees only the applied incident wave). Summing up the contributions from each of them with an appropriate phase factor, the radar cross section of the complete particle is obtained (see Bohren and Huffman 1983):
\begin{equation}
\sigma_r=\left[\frac{36\pi^3}{\rho_{ice}^2}\left|\frac{\epsilon-1}{\epsilon+2}\right|^2\lambda^{-4}\right]m^2f,
\label{sigmar}
\end{equation}
where
\begin{equation}
\label{f}
f=\left[\frac{1}{v}\int_v\exp({2k{\rm i}{\bf r}\cdot{\bf\hat{e}_z}}){\rm d}v\right]^2.
\end{equation}
Here $\rho_{ice}$ and $\epsilon$ are the density and dielectric constant of (solid) ice respectively, $\bf\hat{e}_z$ is the direction of propagation of the incident wave, and $k=2\pi/\lambda$.

 The form factor $f$ represents the deviation from the Rayleigh regime as the overall size of the particle and the wavelength of the incident light become comparable. In the Rayleigh limit where the particle is much smaller than the wavelength, $f=1$, and the radar cross section (\ref{sigmar}) takes the well known $\propto m^2\lambda^{-4}$ dependence. As the particle size increases relative to the wavelength, the form factor falls off with a functional dependance which, in general, depends on the geometry of the particle whose volume is being integrated over in equation \ref{f}. 

The key advantage of the Rayleigh-Gans approach is it allows the scattering cross section for a particle of any shape to be calculated, provided the particle is not too large or too strong a dielectric. Unlike the discrete dipole approximation (Purcell and Pennypacker 1973; Draine and Flatau 1994), the coupling between the elements is neglected, and this places constraints on its applicability. In particular, Bohren and Huffman (1983) quote the conditions $|\sqrt{\epsilon}-1|<1$, $|\sqrt{\epsilon}-1|kr<1$. However, for fractal aggregates (such as our snowflakes), Berry and Percival (1986) have shown that these conditions may be significantly relaxed.
Their calculations show that because of the open structure, multiple scattering between the monomer particles composing fractal aggregates with $d_f\le2$ is negligible, irrespective of the overall size of the aggregate, and for $d_f>2$ only becomes significant when the number monomers per aggregate becomes of order $(kr_0)^{-d_f/(d_f-2)}$. Thus provided that multiple scattering within the monomers themselves is negligible (ie. $|\sqrt{\epsilon}-1|kr_0<1$), which we expect to be reasonable for typical radar frequencies/monomer sizes, the Rayleigh-Gans theory may be safely employed.

The pristine crystal geometry may also play a role in determining the range of applicability of the Rayleigh-Gans theory. Since all interaction between the scatterers (which are assumed to act as equivalent volume Rayleigh spheres) is ignored, the anisotropy of the monomer particles is implicitly neglected. However, Liu and Illingworth (1997) have studied the scattering from a quite elongated hexagonal ice crystal (aspect ratio=3), and found that for $kr_0\ll1$ an equivalent volume sphere gives the same radar cross section to within 1\%. We therefore expect that ignoring the anisotropy of the ice crystals is an acceptable approximation, provided that they are small enough compared to the incident radar wavelength.


	    \bhead{The form factor}          
	    
\begin{figure}                                          
            \centering\includegraphics{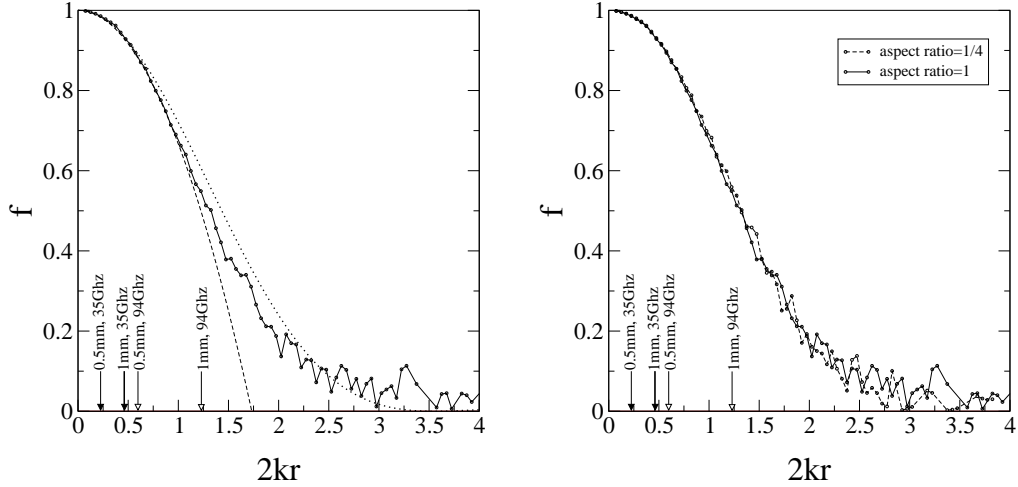}             
            \caption{\label{formfig} The form factor $f$ plotted as a function of $2kr$ for our synthetic aggregates. Left hand panel shows the form factor evaluated for all of the aggregates produced over the course of a simulation, binned and averaged as a function of $2kr$ (solid line with dots). The dashed line shows the small size Guinier approximation $f\simeq1-\frac{1}{3}(2kr)^2$; the dotted line shows the form factor for a sphere with the same radius of gyration $r$ (note $r\simeq0.3D$). The right hand panel highlights the insensitivity of $f$ to pristine particle shape, as the aspect ratio of the pristine columns is varied from unity (solid line) to 0.25 (dashed line) with no appreciable change in the curve.}                                
            \end{figure}

It is apparent from equations \ref{sigmar} and \ref{f} that outside the Rayleigh limit the radar cross section is sensitive to both particle size and shape through the form factor $f$.
For every aggregate produced in our simulations, equation (\ref{f}) was evaluated for a range of wavelengths. The resulting values of $f$ were binned as a function of $(2kr)$ and averaged to yield the curve in figure \ref{formfig}a. 

Given our results on the universality of aggregate snowflake geometry, we anticipate that the form factor will also be independent of the details of the monomer crystals: in figure \ref{formfig}b we illustrate this point by varying the aspect ratio of the pristine crystals and measuring an almost identical form factor. 

We now compare the form factor of our aggregate snowflakes with analytical results from the literature. It is well known (particularly in the polymer physics community) that the asymptotic departure from the Rayleigh limit at small sizes may be approximated as:
\begin{equation}
f\simeq1-\frac{1}{3}(2kr)^2.
\label{guinier}
\end{equation}
This is the Guinier equation (Guinier 1939, Guinier \textit{et al} 1955), and it applies to particles of any shape, provided $2kr$ is small enough. Equation (\ref{guinier}) is plotted alongside the aggregate form factor in figure \ref{formfig}, and proves to be an excellent fit to our simulation data up to $2kr\simeq1$, beyond which it underestimates the aggregate curve.

In meteorological studies, ice particles are often approximated as spheres, in order to utilise Mie's exact solution for the scattering from a dielectric sphere (Mie 1908). The form factor for a sphere is also well known (Bohren and Huffman 1983) and is given by:
\begin{equation}	
f=\left[\frac{3}{u^3}(\sin u-u\cos u)\right]^2
\end{equation}
where $u=2\sqrt{5/3}kr$. Matrosov (1992) has employed this formula as an alternative to the more complex Mie expansion, and showed the results closely mimic the full Mie solution.

Figure \ref{formfig} shows that the correspondance between the spherical model and our aggregates is quite good at small enough sizes, but diverges once $2kr>0.5$. We also note that the sphere model predicts a form factor of zero (ie. no back scatter at all) at $2kr\simeq3.5$, and similar points with $f=0$ at higher values of $2kr$. This undulating form (also found in the exact Mie solution) is a peculiarity of the spherical case, and is not reproduced for our aggregates which do not possess the smooth, symmetrical shape of the idealised dielectric sphere. The errors associated with the spherical approximation are discussed in more detail in section 8.

Values of $2kr$ for typical Cirrus snowflake sizes (0.5 and 1.0mm maximum dimension) at representative radar frequencies (35 and 94 Ghz) are marked on figure \ref{formfig}, and in all cases $2kr<1$. We investigate the consequences of this in the next section, where we show that if the Guinier approximation holds, derivation of average particle size is exceptionally straightforward and may be achieved through a simple analytical formula.

Since radar scattering is rather sensitive to the largest particles in the distribution (even though there may be relatively few of them), we also consider the case of snowflakes beyond the Guinier regime and fit a curve to the complete form factor. We then demonstrate how such a curve may be used to interpret radar data which is outside the Guinier regime.

\ahead{Interpretation of reflectivity data}	
In this section we employ the dynamical scaling property of the distribution, along with the results from the previous section on the scattering from individual snowflakes, in order to study the physical significance of reflectivity measurements and to make use of this understanding to infer the cloud's microphysical properties (average snowflake size, ice water content, etc).

Having calculated the form factor $f$, equation \ref{sigmar} may be evaluated for each snowflake in the scattering volume and the results added, to obtain the reflectivity $\eta=\sum\sigma_r$:
\begin{equation}
\label{eta}
\eta=\left(\frac{c}{\lambda^4}\right)\sum m^2f(2kr)
\end{equation}
where the constant $c=36\pi^3|\frac{\epsilon-1}{\epsilon+2}|^2/\rho_{ice}^2$. Multiple scattering between the snowflakes themselves is assumed to be negligible, given the dilute nature of most ice clouds. Note that the reflectivity is commonly rescaled to give the `radar reflectivity', defined as $Z=\lambda^4\pi^{-5}|\frac{\epsilon-1}{\epsilon+2}|^{-2}\eta$. In what follows we will use $\eta$ for clarity, since its definition follows naturally from the radar cross sections calculated in the previous section. 

\bhead{The Guinier regime (small aggregates)} 

From equation (\ref{eta}), the information contained in measurements of the radar reflectivity is immediately apparent. If the snowflakes are much smaller than the wavelength then $f\simeq1$ and we recover the well known Rayleigh limit $\eta\propto\lambda^{-4}\sum m^2$, and the reflectivity is simply a measure of the second moment of the mass distribution. As the size of the snowflakes and the wavelength become comparable, the particle radius also contributes to the reflectivity through $f$.
Provided the combination of snowflakes size/radar wavelength falls within the Guinier regime $2kr<1$, the form factor is approximated by equation \ref{guinier} and as a result the reflectivity is simply:
\begin{equation}
\label{etag}
\eta=c\left[\frac{\sum m^2}{\lambda^4}-\frac{16\pi^2}{3}\frac{\sum(m^2r^2)}{\lambda^6}\right].
\end{equation}
The first term in the bracket is the Rayleigh result, and for long enough wavelengths (compared to $r$) this behaviour dominates the scattering. As wavelength and particle size become more comparable the second term becomes more significant, and the scattering becomes directly dependent on the particle radius. 
The implication is that if more than one radar wavelength is used, an estimate of average particle radius may be obtained. Given equation \ref{etag}, a natural definition for that average radius is $r_{av}=[\sum m^2r^2/\sum m^2]^{\frac{1}{2}}$, and for two reflectivity measurements $\eta_1$ and $\eta_2$ (at wavelengths $\lambda_1$ and $\lambda_2$) we find:
\begin{equation}
r_{av}=\left[\frac{3}{16\pi^2}\cdot\frac{1-\beta}{1/\lambda_1^2-\beta/\lambda_2^2}\right]^{1/2}
\label{ravg}
\end{equation}
where $\beta=(\eta_1\lambda_1^4)/(\eta_2\lambda_2^4)$, or in terms of radar reflectivities $\beta=Z_1/Z_2$. 

The practical consequence of equation \ref{ravg} is that given two simultaneous measurements at different radar frequencies the average snowflake radius may be calculated \textit{analytically} through equation \ref{ravg}, provided that a) at least one of the radars is operating at a wavelength sufficiently short that some of the snowflakes fall outside the Rayleigh regime, and b) both radars have wavelengths long enough that the snowflakes fall predominantly inside the Guinier regime ($2kr<1$). In previous dual wavelength studies (eg. Matrosov 1998; Hogan \textit{et al} 2000), the relationship between average particle size and $\beta$ has been calculated numerically for spheres and ellipsoids; here we have a simple analytic result, which we expect to be applicable to typical snowflake sizes and radar wavelengths.

Once $r_{av}$ has been calculated then either of the reflectivity measurements may be used to infer $\sum m^2$. Rearranging equation \ref{etag} yields:
\begin{equation}
\label{m2}
\sum m^2=\frac{\eta\lambda^4}{c\left[1-\frac{1}{3}(2kr_{av})^2\right]}.
\end{equation}
In general we would like to use $r_{av}$ and $\sum m^2$ to estimate the ice water content $\mathrm{IWC}=\sum m$. From our definition of the average mass we see that this is simply given by:
\begin{equation}
\mathrm{IWC}=\frac{\sum m^2}{m_{av}}.
\end{equation} 
Thus the IWC may be inferred, provided a measure of average snowflake mass is available. Unfortunately it was demonstrated above that dual wavelength measurements only give a measurement of average radius, not average mass. However equations \ref{fractal} and \ref{ds} provide a means to covert between the two:
\begin{equation}
m_{av}=a\frac{p_2}{p_3}r_{av}^2
\end{equation} 
(taking $d_f=2$). The ratio $p_2/p_3$ has a universal (dimensionless) value, which we measure in our simulations to be around $0.43\pm0.01$; analysis of experimental size distribution data (see appendix B) agrees closely, with a value of $0.44\pm0.01$. The value of the prefactor $a$ by contrast is not universal, depending upon both the size and geometry of the pristine ice crystals making up the snowflakes. For simple, compact crystal habits such as columns or plates there is a linear sensitivity to the crystal size $a\propto r_0$; for dendritic shapes, which have a fractal scaling of their own, we expect the prefactor to be less sensitive, since it is the mismatch between the mass-radius scaling of the crystals and that of the overall aggregate which results in the dependence on crystal size.

\bhead{Larger aggregates}           
	   
For large enough snowflakes relative to radar wavelength, the Guinier formula breaks down, and in this section we identify the point where this occurs, and attempt to fit a curve to describe the scattering at large sizes. In order to achieve this, we rewrite equation \ref{eta} in the form:
\begin{equation}
\eta=\frac{c}{\lambda^4}\left(\sum m^2\right)\mathscr{F}(2kr_{av})
\end{equation}
defining $\mathscr{F}=\sum m^2f/\sum m^2$ (ie. the $m^2$-weighted average of the form factor $f$). We have already identified the form factor for individual aggregates as being a universal function of $2kr$. Because of the dynamical scaling of the distribution and the universal power law scaling between mass and radius, we expect $\mathscr{F}$ be a universal function of $2kr_{av}$. Expressing $\eta$ in this way makes it immediately obvious that only two quantities may be directly derived from the reflectivity: $\sum m^2$ and $r_{av}$. All other quantities (eg. $m_{av}$, IWC) must be inferred from these two, and any other available measurements or assumptions.

\begin{figure}                                          
            \centering\includegraphics{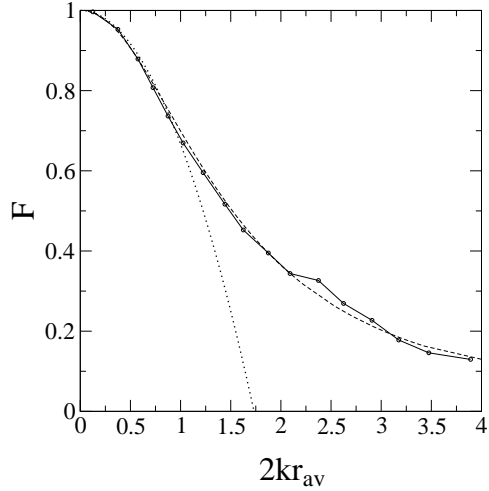}             
            \caption{\label{formd} The $m^2$-weighted average form factor for the whole distribution $\mathscr{F}=\sum m^2f/\sum m^2$, plotted as a function of average size $2kr_{av}$ for our synthetic aggregates (points w/solid line). Dotted line is the Guinier approximation $\mathscr{F}\simeq1-\frac{1}{3}(2kr_{av})^2$; dashed line is our fitted curve (see text for details).}
	    \end{figure}

Figure \ref{formd} shows $\mathscr{F}$ calculated from our simulations. In the Guinier regime, the average form factor is simply $\mathscr{F}\simeq1-\frac{1}{3}(2kr_{av})^2$, and this approximation fits the simulation data well up to $2kr_{av}\simeq1$, beyond which it underestimates the scattering. To try and provide a description for the scattering beyond $2kr_{av}\simeq1$ we have fitted a curve to our simulation data. At small sizes we know that the curve must approach the Guinier formula; at large sizes, it is well known in the physics literature (eg. Viscek 1989) that $\mathscr{F}\propto(2kr_{av})^{-d_f}$. We therefore fit a curve of the form:
\begin{equation}
\label{full}
\mathscr{F}=\frac{1+c_1(2kr_{av})^2}{1+(c_1+\frac{1}{3})(2kr_{av})^2+c_2(2kr_{av})^4}
\end{equation}
which has the correct asymptotics in both limits. As shown in figure \ref{formd}, this provides a good fit to the simulation data with $c_1=12.7$ and $c_2=3.6$.

Given a dual wavelength ratio measurement $\beta$, we may use this fitted curve to estimate the average radius. Noting that $\beta=\mathscr{F}(2k_1r_{av})/\mathscr{F}(2k_2r_{av})$ (where $k_1=2\pi/\lambda_1$, $k_2=2\pi/\lambda_2$) we obtain:
\begin{equation}
\label{cubic}
a'(4r_{av}^2)^3+b'(4r_{av}^2)^2+c'(4r_{av}^2)+d'=0
\end{equation}
where:
\begin{eqnarray}
a'&=&c_1c_2k_1^2k_2^2(\beta k_1^2-k_2^2)\\
b'&=&c_2(\beta k_1^4-k_2^4)\\
c'&=&c_1(\beta k_2^2-k_1^2)+\\
&+&\left(c_1+\textstyle\frac{1}{3}\right)(\beta k_1^2-k_2^2)\\
d'&=&\beta-1.
\end{eqnarray}
Equation \ref{cubic} is a cubic equation in $4r_{av}^2$, and may be solved using standard methods (eg. Press \textit{et al} 1992) to obtain $r_{av}$. The second moment may then be obtained straightforwardly, using either of the reflectivity measurements: $\sum m^2=\eta\lambda^4/c\mathscr{F}(2kr_{av})$. Interpretation of these results in terms of IWC then follows as described in section (a).

\ahead{Inferring other microphysical properties}
Moments of the size distribution other than the ice water content may be also be inferred from reflectivity measurements. In section 5 it was demonstrated that dual wavelength measurements of ice clouds allow the second moment $\sum m^2$ to be measured, along with an average radius $r_{av}$, which may be converted to an average mass $m_{av}$ given knowledge of the prefactor $a$ in the fractal scaling relation. We therefore wish to relate the moments of the size distribution that are of interest to the second moment (which we know). From the dynamical scaling of the distribution (\ref{ds}) it follows that:  
\begin{equation}
\sum m^n=m_{av}^{n-2}\frac{p_n}{p_2}\sum m^2
\end{equation}
for $n\ge1$. Moments with $n<1$ scale differently due to the power law at the small end of the distribution --- see Westbrook \textit{et al} 2004b for details). The ratios $p_n/p_2$ are universal, and may be measured from simulations or derived from experimental size spectra as per Appendix B.

A parameter of significant interest to meteorologists which some authors have sought to derive from radar measurements is the effective radius $r_e$. This characterises the average ratio of particle volume to projected area (Foot 1988):
\begin{equation}
r_e=\frac{3}{2}\frac{\textrm{IWC}/\rho_{ice}}{\sum A}
\end{equation}
where $A$ is the particle projected area. For aggregates however, estimating $r_e$ from the dual wavelength derived average size is not possible. Since the mass is proportional to $r^2$, and likewise for the projected area (the aggregates considered here are fractals with $d_f=2$), the effective radius does not scale with the radius of gyration or maximum dimension; its value is governed by the details of the pristine particles that compose the aggregate snowflakes. Although there is likely to be some correlation between $r_{av}$ and $r_e$ from cloud to cloud, this simply reflects the fact that larger pristine particles tend to yield larger aggregates. Essentially $r_{av}$ is a measure of the overall scale of the aggregate, while $r_e$ is a measure of the size of the pristine crystals. Trying to derive one radius from the other is therefore not possible, and dual wavelength radar measurements do not provide the means to estimate $r_e$.

\ahead{Application to cirrus cloud data set}

\begin{figure}                                          
            \centering\includegraphics{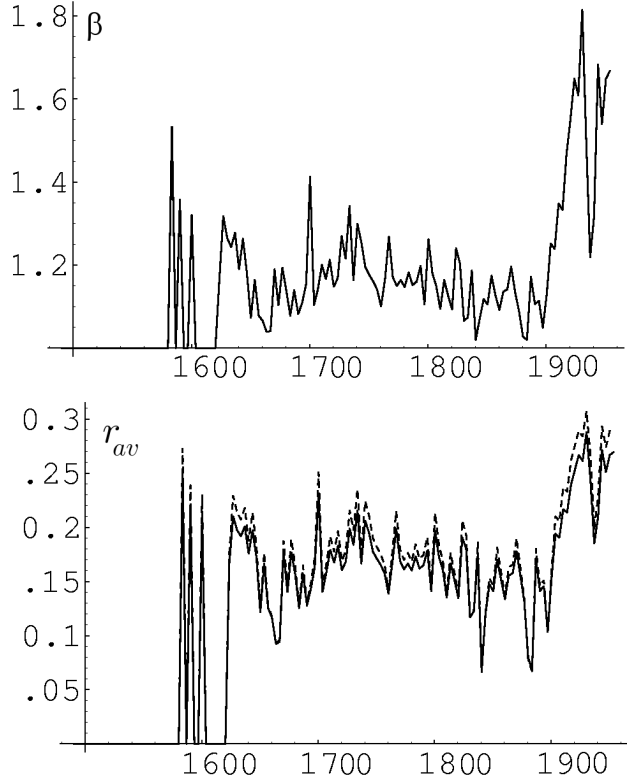}             
            \caption{\label{radarfig} Time series showing the dual wavelength ratio $\beta=Z_{\mathrm{35Ghz}}/Z_{\mathrm{94Ghz}}$ between 15:00 and 19:30 on the 
	    $22^{\mathrm{nd}}$ of June 1996 measured over Chilbolton in the UK (top panel). Shown underneath is the inferred average radius $r_{av}$ derived using both the analytic Guinier formula (dashed line) and the complete fitted curve of section 5b (solid line). Units are in millimetres.}
            \end{figure} 

In this section we apply the two methods for estimating $r_{av}$ to a dual wavelength dataset from a real ice cloud. On the $22^{\mathrm{nd}}$ of June 1996, vertically pointing radar measurements of a cirrus cloud over Chilbolton in the UK were made using 35 and 94Ghz radars. This data was presented in Hogan \textit{et al} (2000), and the full experimental details are given in that paper. Here we analyse the radar reflectivity time series from one representative altitude ($\sim$ 6 km) where the data shows approximately the full range of reflectivity values measured during the experiment. From these measurements we have calculated the dual wavelength ratio $\beta=Z_{\mathrm{35Ghz}}/Z_{\mathrm{94Ghz}}$ and this is plotted as a function of time in figure \ref{radarfig}. From this we have used first the Guinier approximation (small flakes, as described in section 5a) and then our fitted curve (section 5b) to infer the average radius $r_{av}$. These results are also plotted in figure \ref{radarfig} and 
it is immediately apparent that there is very little to distinguish the Guinier-derived radius from the one obtained using the full fitted curve. The implication is that the snowflakes are small enough that they lie within the Guinier regime, and since the average radius $r_{av}$ only reaches around 0.3mm at most ($\sim$ equivalent to 1mm maximum dimension), this means $2kr_{av}\le1.2$ for 94Ghz, and $2kr_{av}\le0.4$ for 35Ghz. Referring to section 5 these values both fall roughly within the Guinier regime.

\begin{figure}                                          
            \centering\scalebox{0.8}[0.8]{\includegraphics{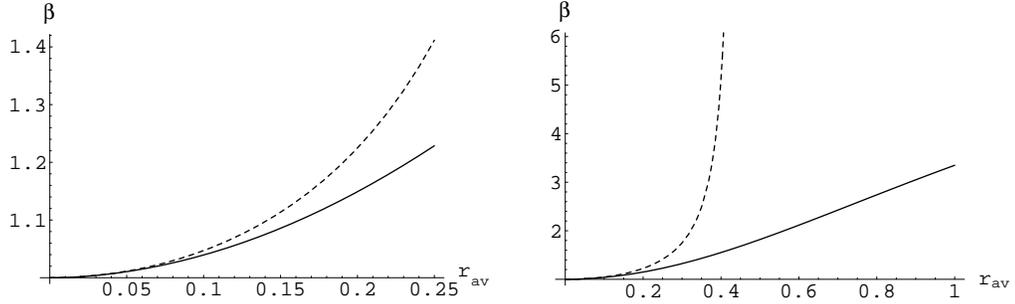}}             
            \caption{\label{fg3594comp} Dual wavelength ratio $\beta$ as a function of average radius $r_{av}$ in millimetres for 35- and 94-GHz radars. Black line shows the prediction of our fitted curve (\ref{full}); dashed line shows the Guinier approximation. Left hand panel shows good agreement between the Guinier approximation and our fitted curve at small sizes ($r<0.25$); the right hand panel shows the behaviour at larger sizes, where there is a marked divergence.}
\end{figure}

For the studied cirrus cloud at least, the Guinier formula is a good approximation. We note however that even here the largest particles were on the edge of the regime ($2kr_{av}\simeq1$); for thicker clouds containing larger aggregates this formula may not be sufficient, and the complete fitted curve (\ref{full}) should be used. Figure \ref{fg3594comp} shows a plot of dual wavelength ratio as a function of average radius for 35- and 94-Ghz radars, as predicted using both (\ref{full}) and the Guinier approximation. It is apparent that the Guinier formula holds up well to around $r_{av}\simeq0.3$mm, beyond which point it rapidly breaks down. 
It is useful to see however, that at least for parts of the cloud dominated by smaller aggregates, the Guinier approximation does hold and that a simple analytic expression provides an accurate estimate for $r_{av}$. In both cases there is a need for the methods to be validated through simultaneous aircraft measurement of particle mass and radius distributions. 

\ahead{Comparison with spherical models}
Many authors have used a spherical approximation when estimating the back scatter from snowflakes, modelling them as homogeneous mixtures of air and ice. The motivation for this kind of approximation is that the scattering from a sphere with a given diameter and dielectric constant may be calculated exactly using Mie theory. Here we consider how closely the spherical model matches the results from our synthetic aggregates.

To mimic the results for the radar cross section by our simulated aggregates using a sphere, we attempt to match up both the form factor $f$ for the aggregates, and the $m^2$ dependance for $2kr\ll1$.
In figure \ref{formfig} the form factor for a sphere was plotted alongside that for our simulated aggregates. There is good agreement up to $2kr\simeq0.5$, after which the curves diverge somewhat. Thus for small aggregates at least, an equivalent sphere is an accurate representation provided that the mass and radius of gyration are conserved.
We focus therefore on this regime, and consider the errors at larger sizes later. In order to match up the scattering from our sphere with that of the aggregate it is intended to model, we must ensure the radius of gyration of the sphere is the same as that of the aggregate. For our simulated aggregates, the radius of gyration and maximum span are found to scale linearly with $r\simeq0.3D$, whereas a sphere of diameter $D_{sp}$ has $r=0.39D_{sp}$ (see Appendix A). To match these up then requires that the diameter of the equivalent sphere be $D_{sp}=0.77D$. This ensures that the sphere has the same radius of gyration as the aggregate it is intended to represent, and hence the form factor shown in figure \ref{formfig}. A different choice of $D_{sp}$ would result in the curve being squashed or stretched along the $2kr$ axis, and under/over-estimating the back scatter for $2kr<0.5$.

\begin{figure}
\centering\scalebox{0.8}[0.8]{\includegraphics{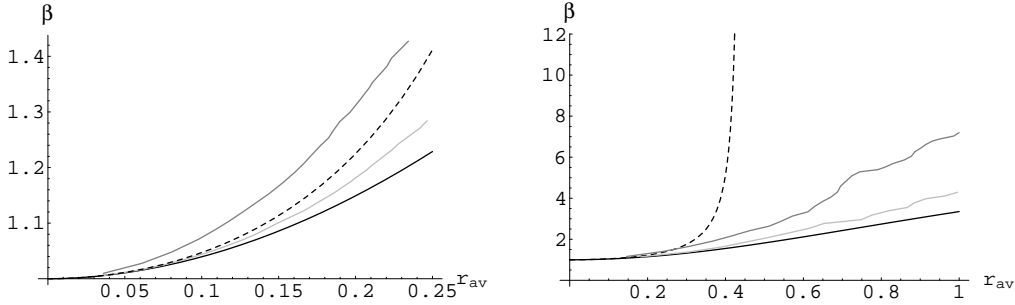}}
\caption{\label{betarav3594} Dual wavelength ratio $\beta$ as a function of $r_{av}$ for 35- and 94-GHz radars as per figure \ref{fg3594comp}. Black line shows the prediction of our fitted curve (\ref{full}); dashed is the Guinier approximation. The grey lines show measurements of $\beta$ from a computer simulation in which a sphere was used to model the aggregates, where: (i) the radius of gyration of the aggregate and the sphere used to model it were matched (light grey), and (ii) the diameter of the sphere was chosen to be the maximum dimension of the aggregate (dark grey).}
\end{figure}

To estimate the error involved in this approach we have run computer simulations where we calculate the dual wavelength ratio at 35- and 94-Ghz, modelling the scattering from our aggregates as that of spheres. We have considered two cases: (i) where we choose each sphere's diameter so as to match the radius of gyration of the sphere with that of the aggregate as discussed above, and (ii) where we choose the diameter of the sphere to match the maximum dimension of the aggregate $D$ (a model suggested by some authors in the literature): the results of these simulations are shown in figure \ref{betarav3594}.
When the radii of gyration are matched, the agreement at small sizes is good, with only around a 5\% error in $\beta$ relative to our fitted curve at $r_{av}=0.25$mm. At larger sizes the divergence is more apparent, and the sphere model overestimates $\beta$ by around 25\% at $r_{av}=1$mm.
Model (ii) performs rather less well: matching the diameter of the sphere to the maximum dimension, the dual wavelength ratio is overestimated by around 20\% at $r_{av}=0.25$mm, and by more than 100\% at $r_{av}=1$mm. We conclude that both methods overestimate $\beta$ for a given $r_{av}$, but matching the radius of gyration of the sphere with the aggregate results in much lower errors than matching their maximum dimension.

\begin{figure}
\centering\scalebox{1.0}[1.0]{\includegraphics{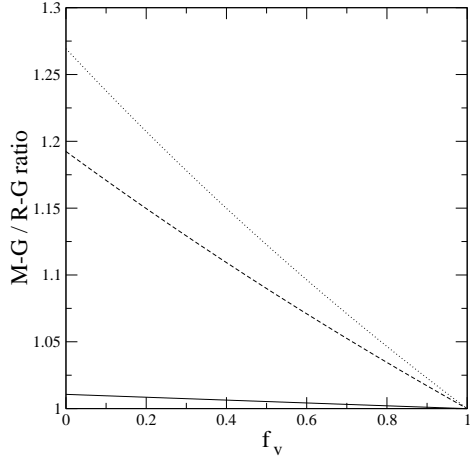}}
\caption{\label{goodness}Ratio comparing the back scatter cross section calculated using the Maxwell-Garnett sphere model, and the Rayleigh-Gans theory, for $2kr\ll1$ (the Rayleigh limit). Solid line represents spherical inclusions ($\kappa=0.58$), dashed is Meneghini and Liao ($\kappa=0.63$), and dotted is rod shaped inclusions ($\kappa=0.65$).}
\end{figure}

The second dependence that must be matched is the snowflake mass. Since the snowflakes are not solid ice spheres, information about how much of the mixture is air and how much is ice must be introduced by using an `average' dielectric constant $\epsilon_{av}$ weighted by the volume fraction of ice in the sphere $f_v=m/\frac{\pi}{6}(D_{sp})^3\rho_{ice}$. The most common prescription for $\epsilon_{av}$ is the Maxwell-Garnett (1904) formula:
\begin{equation}
\label{mg}
\epsilon_{av}=\frac{1-f_v+\epsilon\zeta f_v}{1-f_v+\zeta f_v}
\end{equation}
where the parameter $\zeta$ reflects the distribution of sizes and shapes of the ice `inclusions' in the mixture (see Bohren and Huffman 1983 for more details). For spherical inclusions $\zeta=0.58$.
To check that introducing the aggregate density in this way yields the same $m^2$ dependance as per equation \ref{sigmar}, we study the Rayleigh limit where the back scatter cross section of our equivalent sphere is given by:
\begin{equation}
\label{rl}
\sigma_r=\frac{\pi^5}{\lambda^4}\left|\frac{\epsilon_{av}-1}{\epsilon_{av}+2}\right|^2D_{sp}^6.
\end{equation}
For solid ice spheres $\epsilon_{av}=\epsilon$, and since $m=\frac{\pi}{6}(D_{sp})^3\rho_{ice}$, one recovers equation \ref{sigmar} with $f=1$. For small volume fractions (the case most relevant to fractal aggregates, since the overall density is proportional to $r^{d_f-3}$) we calculate the ratio of the back scatter obtained from equation \ref{rl} to that obtained from equation \ref{sigmar} as a function of the volume fraction $f_v$: in the Rayleigh limit this turns out to be simply $|\frac{\epsilon_{av}-1}{\epsilon_{av}+2}|^2/\{|\frac{\epsilon-1}{\epsilon+2}|^2f_v^2\}$. This ratio is unity at $f_v=1$; even as $f_v\rightarrow0$ the error is only around $1\%$ (see figure \ref{goodness}). 
Other values of the parameter $\zeta$ are also in common usage - the assumption of needle shape inclusions gives $\zeta=0.65$ and Meneghini and Liao (1996) have proposed a value of $\zeta=0.63$. These alternate values of $\zeta$ essentially represent the polarisability of the inclusions (which we interpret in the case of our aggregates as being the monomer particles), and they predict a back scatter increased by as much as 25\% as $f_v\rightarrow0$, as shown in figure \ref{goodness}. In section 4 it was commented that in the Rayleigh limit the scattering from the individual monomers should be well represented by a sphere of the same volume; if this is the case then the Rayleigh-Gans approximation ought to be accurate, and these two values of $\zeta$ overestimate the scattering quite significantly, particularly at low volume fractions. This is perhaps a reflection of the fact that `effective medium' theories such as the Maxwell-Garnett rule are intended for strong dielectrics at relatively large volume fractions, neither of which is really the case for ice aggregates.

In summary, the spherical approximation matches the results for our aggregate snowflakes provided that i) the diameter is chosen so as to match the radius of gyration of the sphere to the snowflake it is intended to model, and ii) the parameter $\zeta$ is chosen so as to give the correct $m^2$ dependence in the Rayleigh limit. If the size of the aggregates is not too close to the wavelength of the radar, then the error in $\beta$ is only $5\%$ or so; for larger aggregates this may increase to 25\% or more. It is important to note that a 5\% error in $\beta$ may translate into a rather more significant error in the derived parameters such as $r_{av}$ and IWC.

\ahead{Conclusions}
The radar cross sections for our simulated snowflakes have been calculated using the Rayleigh-Gans theory by treating the aggregates as being an assembly of independant Rayleigh scatterers, and summing up the contributions from each element in that assembly. In this theory, the scattering is proportional to the square of the particle mass (as in the Rayleigh limit) and the form factor $f$, a dimensionless function which describes the depedence on particle size and geometry. This function is universal (ie. independent of the size and shape of the monomer particles), and $f$ has been calculated for our simulated aggregates. 

This is (to the authors' knowledge) the first time that the scattering from realistic aggregate snowflakes has been calclulated. It also appears to be the first time the Rayleigh-Gans theory has been employed to study radar scattering by ice particles, with the exception of Matrosov (1992) who used it only as an approximation to the exact Mie solution for a sphere. Ideally we would like to verify the results obtained in this paper using the discrete dipole approximation which accounts for the interaction between the scattering elements. However, this places heavy demands on computer time and memory, since a large number of dipoles are needed to accurately model the detailed structure of the aggregates, and calulations for many possible realisations and orientations must be done in order to obtain good statistics. So far this has not been achieved. Also, in order to test the accuracy of the theoretical results in sections four and five, simultaneous radar and aircraft observations of ice clouds are needed. 

The form factor for our simulated snowflakes was compared to analytical results for $f$ from the scattering literature, the most successful of which in describing the data was the Guinier result $f=1-\frac{1}{3}(2kr)^2$, which fits our data well up to $2kr\simeq1$. The result for the sphere is a good approximation below $2kr\simeq0.5$, beyond the curves diverge somewhat, the sphere first over- then under-estimating the back scatter. 

Having calculated the scattering cross section of individual aggregates, these were then integrated over the size distribution, and the results interpreted in terms of moments of that distribution. The reflectivity is proportional to the second moment of the mass distribution $\sum m^2$ and the average form factor $\mathscr{F}=\sum m^2f/\sum m^2$. We have calculated $\mathscr{F}$ as a function of the average radius $r_{av}$, and fitted a curve to our simulation data.
The application of our results to the inference of average snowflake size from radar data has been investigated, and for snowflakes sufficiently small compared to the wavelength, the Guinier approximation provides a very simple analytic expression for $r_{av}$ in terms of the dual wavelength ratio $\beta$. At larger values of $2kr_{av}$, the Guinier curve underestimates the back scatter: we therefore fitted a curve to our simulation data and used that curve to provide a method to interpet the dual wavelength data.

Ideally, we would like to interpret the radar reflectances in terms of the ice water content (total mass per unit volume) in the cloud. Since reflectivity is proportional to $\sum m^2$, some measure of average particle mass is required. We have shown that it is possible to derive the average particle radius $r_{av}$ from dual wavelength radar data and this may be converted to an average mass, provided that some measurement or prescription for the prefactor $a$ in the fractal scaling relation is available: $a$ is a function of the pristine crystal geometry and size.

Some authors have attempted to derive the effective radius (characterising the ratio of particle volume to projected area) from radar measurements. However we have shown that $r_e$ depends only on the monomer particles (to which the radar is insensitive) and not the overall size of the aggregate as characterised by $r_{av}$. Direct inferral of $r_e$ then is not possible from dual wavelength radar data.

Finally, the errors involved in modelling our simulated snowflakes as air/ice spheres using the Maxwell-Garnett mixture theory were analysed. It was found that provided some care was taken in constructing the equivalent sphere, and the radius is not too large, the error is around 5\% ($r_{av}=0.25$mm with 35/94Ghz radars), increasing to around 25\% at larger sizes ($r_{av}=1$mm). It is worth noting that such an error in the dual wavelength ratio has the potential to translate into a much larger error in the derived parameters, especially if $\beta$ is close to unity.

\acks

This work was funded by the Engineering and Physical Sciences Research Council and the Met Office. We would like to thank Robin Hogan (Reading University) for supplying the dual wavelength Cirrus data.

\appendix
\abheadx{Appendix A}{The radius of gyration}

In this paper we have used the radius of gyration $r$ as the characteristic aggregate length scale. It is defined by splitting the particle into small volume elements, with position $\mathbf{r}$ and mass $\mathrm{d}m$. Then:
\begin{equation}
r=\left[\frac{\int|\mathbf{r}|^2\mathrm{d}m}{\int\mathrm{d}m}\right]^2,
\end{equation}
integrating over the particle volume. This radius is a natural length scale to use for the scattering calculations because it is closely related to the integral for $f$ in equation 6 at small $k\mathbf{r}$, leading to the Guinier expansion $f\simeq1-\frac{1}{3}(2kr)^2$. For our aggregates it is linearly related to most other characteristic length scales, in particular the maximum dimension $D$ of the snowflakes ($r\simeq0.3D$). 
				       %

\abheadx{Appendix B}{Inferring the universal moment ratios $p_n/p_2$ from experimental snowflake span distributions}

Here we show how the universal ratios $p_2/p_3$ (used to link $m_{av}$ to $r_{av}$) and $p_n/p_2$ (used to convert $\sum m^2$ and $m_{av}$ into a different moment of the distribution $\sum m^n$) may be calculated from experimental snowflake span distributions. Using equation \ref{fractal} the moments of the radius distribution $\mathscr{M}_n$ are given by:
\begin{equation}
\mathscr{M}_n=\int_0^{\infty}(m/a)^{n/d_f}n(m,t)dm
\end{equation}
which, using the dynamical scaling property (\ref{ds}), and taking $d_f=2$ is:
\begin{equation}
\mathscr{M}_n=(m_{av}/a)^{n/2-\xi}p_{n/2}
\label{nnn}
\end{equation}
From the definition of $m_{av}$, one has $p_2/p_1=1$. We therefore obtain the ratios $\mathscr{M}_4/\mathscr{M}_2=m_{av}/a$, and $\mathscr{M}_6/\mathscr{M}_4=(m_{av}/a)(p_3/p_2)$. The ratio we seek to calculate, $p_2/p_3$, is then simply:
\begin{equation}
\label{p2p3}
\frac{p_2}{p_3}=\frac{\mathscr{M}_4^2}{\mathscr{M}_2\mathscr{M}_6}.
\end{equation}
We note that it makes not difference whether we use moments of the radius distribution ($r$) or of the span distribution ($D$) since the end result is dimensionless, and $r\propto D$. We have calculated $p_2/p_3$ from the experimental particle span distributions presented in Westbrook \textit{et al} (2004a,b), and the ratio is found to be $0.44\pm0.02$.

The ratio $p_n/p_2$ may be obtained in a similar fashion. From equation \ref{nnn} above:
\begin{equation}
\frac{\mathscr{M}_{2n}}{\mathscr{M}_{4}}=\left(\frac{m_{av}}{a}\right)^{n-2}\frac{p_n}{p_2}
\end{equation}
and since $\mathscr{M}_4/\mathscr{M}_2=m_{av}/a$, the ratio we want is simply:
\begin{equation}
\frac{p_n}{p_2}=\frac{\mathscr{M}_{2n}\mathscr{M}_2^{n-2}}{M_4^{n-1}}.
\end{equation}


\references

\bib{Bohren,~C.~F. and Huffman,~D.~R.}{1983}{{\em Absorption and scattering of light by small particles.} 
John Wiley and Sons, New York}
	    
\bib{Berry,~M.~V. and Percival,~I.~C.}{1986}{Optics of fractal clusters such as smoke. {\em Opt. Acta,} 
{\bf 33,} 577--591}

\bib{Draine,~B.~T. and Flatau,~P.~J.}{1994}{Discrete dipole approximation and its application for scattering calculations. {\em J. Opt. Soc. Am. A,} 
{\bf 11,} 1491--1499}

\bib{Heymsfield,~A.~J., Lewis,~S., Bansemer,~A., Iaquinta,~J., Miloshevich,~L.~M., Kajikawa,~M., Twohy,~C. and Poellot,~M.~R.}{2002}{A general approach for deriving the properties of cirrus and stratiform ice cloud particles.. {\em J. Atmos. Sci.,} 
{\bf 59,} 3--29}

\bib{Hogan,~R.~J., Illingworth,~A.~J. and Sauvageot,~H.}{2000}{Measuring crystal size in cirrus using 35- and 94-Ghz radars. {\em J. Atmos. \& Ocean. Tech.,} 
{\bf 17,} 27--37}

\bib{Locatelli,~J.~D. and Hobbs,~P.~V.}{1974}{Fall speeds and masses of solid precipitation particles. {\em J. Geophys. Res.,} 
{\bf 79,} 2185--2197}

\bib{Matrosov,~S.~Y.}{1992}{Radar reflectivity in snowfall. {\em IEEE. Trans. Geosci. \& Rem. Sens.,} 
{\bf 30,} 454--461}

\bib{}{1998}{A dual-wavelength radar method to measure snowfall rate. {\em J. Appl. Met.,} 
{\bf 37,} 1510--1521}

\bib{Mitchell,~D.~M.}{1996}{Use of mass- and area- dimensional power laws for determinating precipitation particle terminal velocities. {\em J. Atmos. Sci.,} 
{\bf 53,} 1710--1723}

\bib{Purcell,~E.~M. and Pennypacker,~C.~R.}{1973}{Scattering and absorption of light by nonspherical grains. {\em Astrophys. J.,} 
{\bf 186,} 705--715}

\bib{Westbrook,~C.~D., Ball,~R.~C., Field,~P.~R. and Heymsfield,~A.~J.}{2004a}{Universality in snowflake aggregation. {\em Geophys. Res. Lett.,} 
{\bf 31,} L15104--15107}

\bib{}{2004b}{A theory of growth by differential sedimentation, with application to snowflake formation. {\em Phys. Rev. E,} 
{\bf 70,} 021403}

\end{document}